\documentclass[twocolumn,twoside,slac]{revtex4}
\usepackage{graphicx}
\usepackage{fancyhdr}
\begin{document}

\title{A data Grid testbed environment in Gigabit WAN with HPSS }


\author{Atsushi Manabe, Setsuya Kawabata, Youhei Morita, Takashi Sasaki, Hiroyuki Sato, Yoshiyuki Watase and Shigeo Yashiro}
\affiliation{High Energy Accelerator Research Organization,KEK, Tsukuba,Japan} 
\author{Tetsuro Mashimo,Hiroshi Matsumoto, Hiroshi Sakamoto,Junichi Tanaka, Ikuo Ueda}
\affiliation{International Center for Elementary Particle Physics
University of Tokyo (ICEPP, Tokyo, Japan)}
\author{Kohki Ishikawa, Yoshihiko Itoh, Satomi Yamamoto }
\affiliation{ IBM Japan, Ltd. }
\author{Tsutomu Miyashita }
\affiliation{KSK-ALPA co. ltd.}
\begin{abstract}
For data analysis of large-scale experiments such as LHC Atlas and
other Japanese high energy and nuclear physics projects, we have
constructed a Grid test bed at ICEPP and KEK. These institutes are
connected to national scientific gigabit network backbone called
SuperSINET. In our test bed, we have installed NorduGrid middleware
based on Globus, and connected 120TB HPSS at KEK as a large scale
data store. Atlas simulation data at ICEPP has been transferred and
accessed using SuperSINET. We have tested various performances and
characteristics of HPSS through this high speed WAN. 
The measurement includes data access perforance comparison between
connections with low latency LAN and long distant WAN.
\end{abstract}

\maketitle

\thispagestyle{fancy}


\section{Introduction}

For the Atlas Japan collaboration, 
International Center for Elementary Particle Physics, 
University of Tokyo(ICEPP) will build a ``Tier-1'' regional
center
and  High Energy Accelerator Research
Organization(KEK) will  build a  ``Tier-2'' regional center 
for the Atlas experiment of the Large Hadron Collider (LHC) project at
CERN.
The two institutes are connected by the SuperSINET which 
is an ultrahigh-speed network for Japanese academic researches.
On the network a Grid test bed  was
constructed to study requisite functionality and performance issues 
for the tiered regional centers.

High Performance Storage System (HPSS) with high density digital tape
libraries could be a key component to handle petabytes of data
produced by Atlas experiment and to share such data among the regional
collaborators. HPSS parallel and concurrency data transfer mechanisms,
which support disk, tape and tape libraries, are effective and scalable
to support huge data storage. This paper describes about integration
of HPSS into a Grid architecture and the performance measurement of
HPSS in use over a high-speed WAN.

\section{Test bed system}

\begin{figure}
\includegraphics[width=65mm]{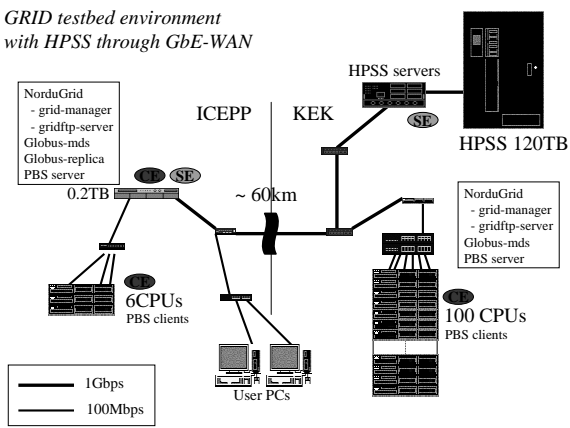}
\vspace{1cm}
\caption{Layout of the test bed hardware}
\label{hard_conf}
\end{figure}

The computer resources for the test bed were installed 
in ICEPP and KEK site. One Grid server in each site and HPSS servers in KEK were 
 connected with 1-Gbps Ethernet through the SuperSINET.
All resources including network 
were isolated from other users and dedicated for the test.
Figure~\ref{hard_conf} and Table~\ref{hard_conf_tab} shows our
hardware setup.  

\begin{figure}[h]
\includegraphics[width=65mm]{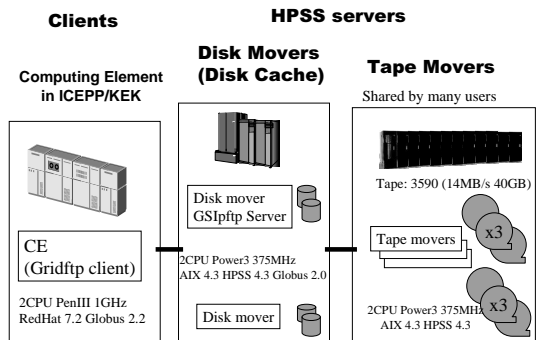}
\vspace{0.5cm}
\caption{HPSS players.}
\label{hpss}
\end{figure}

Three storage system components were employed. A disk storage server
shared its host with the Grid server each  at KEK and ICEPP. The
remaining HPSS software components were used in the KEK Central
Computer system. The HPSS data flow is depicted in Fig.~\ref{hpss}.
The HPSS Servers including core servers, disk movers, and tape movers 
are tightly coupled by an IBM SP2 cluster network switch.

In the case of original pftp(parallel ftp with Kerberos
authentication) server performance
measurement, pftpd was run in the core HPSS server. In the case of
GSI-enabled HPSS server which will be mentioned in the section 
\ref{GSI-enabled-pftp},
 pftpd was run in the same processors with the
disk mover. The disk movers were directly connected to the test bed
LAN through their network interface cards. Two HPSS disk movers were
dedicated to the test.

NorduGrid middleware ran on the Grid servers.  Other computing
elements (CE) acted as a Portable Batch System (PBS)~\cite{PBS}
 that was not required to install with the NorduGrid middleware.

\begin{table}[t]
\begin{center}
\caption{Test bed Hardware}
\begin{tabular}{|l|l|}

\hline
 ICEPP 	& {\bf  Grid and PBS server 		}	\\
       	&  1 $\times$ Athlon 1.7GHz 2CPU  	\\	
       	& {\bf Computing Element 	}		\\
       	&  4 $\times$ pentium III 1.4GHz  2CPU 	\\
\hline
 KEK  	& {\bf Grid and PBS server }			\\
 	& 1 $\times$ Pentium III 1GHz 2CPU 	\\
	& {\bf Computing Element }			\\
 	&  50 $\times$ pentium III 1GHz  2CPU 	\\
 	& {\bf HPSS disk mover }			\\
 	&  2 $\times$ Power3 375MHz 		\\
 	& {\bf HPSS tape mover and Library	}	\\
 	&  19 $\times$ Power3 375MHz, IBM 3590 	\\
\hline
\end{tabular}
\label{hard_conf_tab}
\end{center}
\end{table}

The NorduGrid\cite{nordugrid} is a pioneer Grid project in Scandinavia that added
upper layer functionality, which is necessary to HEP computing, on the
Globus tool kit. The middleware was simple to understand and offered
functionality sufficient for our test bed study.

Table~\ref{soft_conf_tab} shows the versions of middleware used in the
test bed.

\begin{table}[t]
\begin{center}
\caption{Test bed Software}
\begin{tabular}{|l|l|}

\hline
  software     	& version 	\\
\hline
  Globus 	& 2.2.2 	\\
  NorduGrid 	& 0.3.12 	\\ 
  PBS	 	& 2.3.16 	\\ 
  HPSS		& 4.3		\\	
\hline
\end{tabular}
\label{soft_conf_tab}
\end{center}
\end{table}

\section{HPSS performance over high-speed WAN}

\subsection{Basic network performance}

\begin{figure}[h]
\includegraphics[width=65mm]{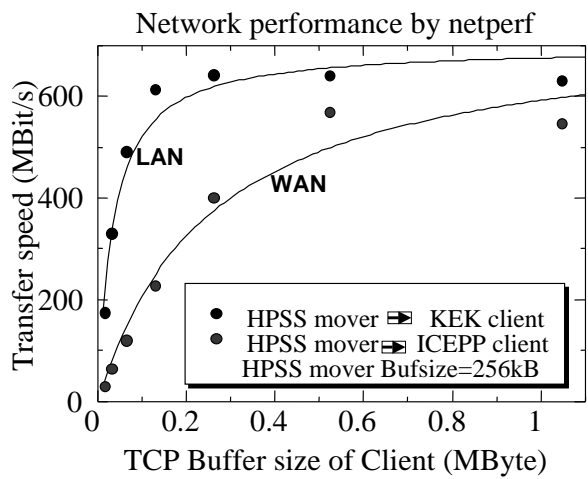}
\vspace{0.5cm}
\caption{Basic GbE network transfer speed measured by netperf.}
\label{basic_net}
\end{figure}

Before end-to-end measurement, basic Gigabit Ethernet performance
between IBM HPSS servers and a host at ICEPP through the WAN
as well as a host on the KEK LAN was measured using netperf
\cite{netperf}.
It is shown in Figure \ref{basic_net} as a function of the TCP buffer
size of the client. Round Trip Time (RTT) averaged was 
3 to 4 ms. The network quality of service was quite good and almost free from
packet loss ($ < 0.1 \%$). In this measurement, maximum TCP window size in
HPSS server had 256kB (the buffer size of 256kB optimized to IBM SP2
switching network). The clients at both KEK  and ICEPP
had 64MB. Due to rather slower clock-speed proceessors on the HPSS servers
the maximum raw TCP transfer performance was limited  below
1Gbps.
As seen in the graph, network access performance through both LAN and
WAN became almost equivalent and saturated beyond 0.5MB buffer size.

\begin{figure}[h]
\includegraphics[width=65mm]{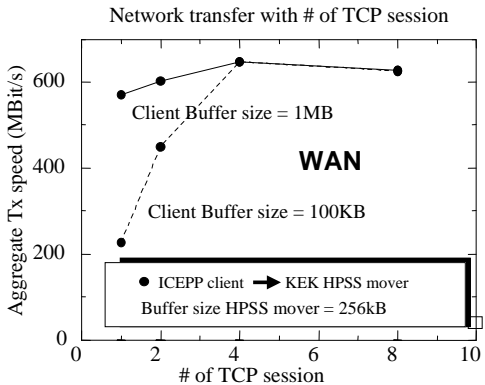}
\caption{Network performance with no. of TCP stream sessions measured
  by netperf.}
\label{basic_net_ses}
\end{figure}

Figure~\ref{basic_net_ses} shows the network performance with the number of
simultaneous transfer sessions through the WAN.
In the situation where TCP buffer size was 100KB,  
up to 4 parallel simultaneous stream sessions improved network
throughput.
Using greater buffer size than 1MB, multiple stream sessions did not
improve the aggregate network transfer speed. 

\subsection{HPSS client API performance }

\begin{figure}[h]
\includegraphics[width=65mm]{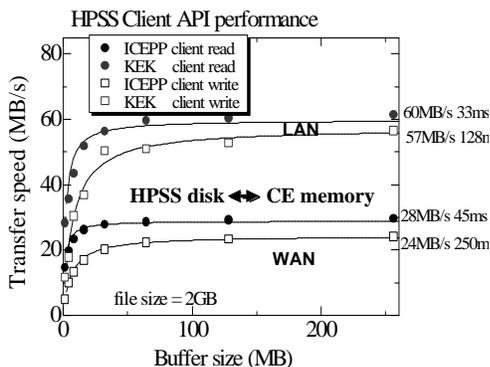}
\caption{HPSS client API performance }
\label{HPSS_client_API}
\end{figure}

Figure~\ref{HPSS_client_API} shows data transfer speed by using  the HPSS
client API and comparison between access from LAN and over WAN.
 The transfer was from/to the disk of HPSS disk-mover disk to/from
client host memory. 
The transferred file size was 2GB in all case.
Disk access speed  in the disk mover was 80MB/s.
It shows that even with a larger API buffer size in the client API,
WAN access speed was about a half of LAN access both for reading and
writing from/to HPSS server. 

To increase HPSS WAN performance in future tests, the newer pdata
protocol provided in HPSS 4.3 can be employed. This will improve pget
performance. To get the same effect on pputs, the pdata-push protocol
provided in HPSS 5.1 is required.

The existing mover and pdata protocols are driven by the HPSS mover
with the mover requesting each data packet by sending a pdata header
to the client mover.  The client mover then sends the data.  This
exchange creates latency on a WAN. The pdata-push protocol allows the
client mover to determine the HPSS movers that will be the target of
all data packets when the data transfer is set up.   This protocol
eliminates the pdata header interchange and allows the client to just
flush data buffers to the appropriate mover.  The result is that the
data is streamed to the HPSS mover by TCP at whatever rates it can be
delivered by the client side mover and written to the HPSS mover
devices.

\subsection{pftp client - pftpd server transfer speed }

\begin{figure}[h]
\includegraphics[width=80mm]{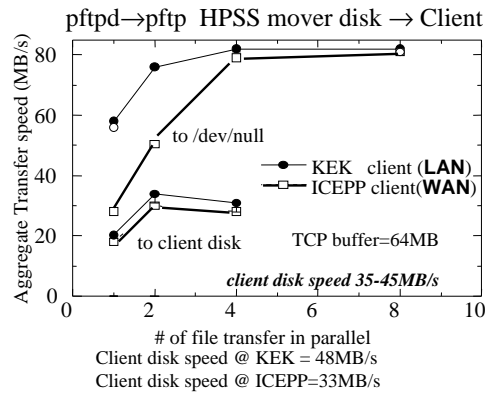}
\caption{Performance pftpd-pftp client read to client /dev/null and disk}
\label{pftp_readdisk}
\label{pftp_devnull}
\end{figure}

Figure~\ref{pftp_devnull} shows data transfer speed by using  HPSS
pftp from HPSS disk mover to client /dev/null dummy device.
Again as in the previous HPSS client API transfer, even with  a
pftp buffer size of 64MB, access speed from WAN was about a half of LAN access.
In addition, enabling single file transfer with multiple TCP stream
by using the pftp `pwidth' option was not effective in our situation.
In our server layout, two disk mover hosts each had two RAID disks.
Therefore, up to 4 concurrent file transfers could effect higher
network
utilization and overall throughput, and was so seen in WAN and LAN
access case.
In the same figure (Fig.~\ref{pftp_readdisk}) data transfer speed was 
shown from HPSS disk mover to the client disks which had writing performance of
35-45MB/s. Though disks both in server and client hosts had the access
speed exceeding 30MB/s and also network transfer speed exceeded
80MB/s, overall transfer speed  dropped to 20MB/s.
It is because these three resources were not accessed in parallel but in series.

\begin{figure} [h]
\includegraphics[width=80mm]{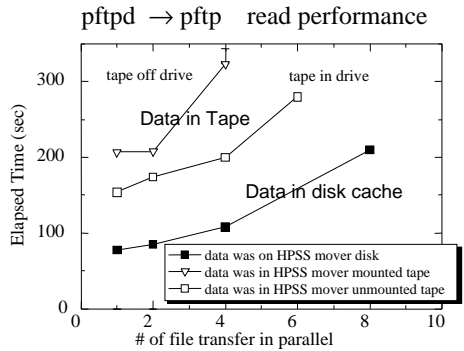}
\caption{pftpd-pftp read to client disk from tape archive performance}
\label{HPSS_tape}
\end{figure}

Figure~\ref{HPSS_tape} shows elapsed time for accessing data in tape
library. Thanks to HPSS functionality and an adequate number of
tape movers and tape drives, the data I/O throughput 
 scaled with the number of concurrent file transfers. 
However, since the library had only two accessors and could load upto two
tape cassetes to drives simultaneously, in the case where data in more than three 
off-drive tapes is required to access, the throughput goes down.

\begin{figure}[h]
\includegraphics[width=85mm]{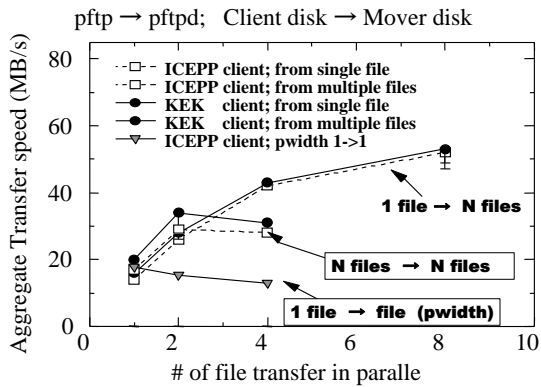}
\vspace{0.5cm}
\caption{pftpd-pftp write to server cache-disk performance}
\label{pftp_writedisk}
\end{figure}

Comparison of writing to HPSS disk mover
 from client  over WAN and LAN is shown in Fig.~\ref{pftp_writedisk}.
In the figure, `N files $\to$ N files', for example, means that `reading' N files
simultaneously at client and `writing' N files to the server.
In our setup, HPSS server had 4 independent disks but client had only one.
Reading multiple files in parallel from a single disk at client side
degrades the aggregate performance by contention of disk heads.

%\begin{figure}[h]
%\includegraphics[width=65mm]{g4_2.eps}
%\vspace{0.5cm}
%\caption{pftpd-pftp read to client /dev/null performance}
%\label{pftp_devnull}
%\end{figure}

\section {GSI-enabled pftp }
\label{GSI-enabled-pftp}
GridFTP\cite{gridftp} is a standard protocol for building data GRID 
and supports the featues of 
Grid Security Infrastructure (GSI),
multiple data channels for parallel transfers, partial file transfers,
third-party transfer and  reusable and authenticated data channels.

The pftp and ftp provided with HPSS software was not required or
designed to support data Grid infrastructure. For future releases,
HPSS Collaboration Members have introduced data Grid pftp requirements
and the HPSS Technical Committee (TC) has convened a Grid Working
Group to propose a development plan. As an interim and partial HPSS
data Grid interface solution, the HPSS Collaboration is distributing
the GSI-enabled pftp developed by Lawrence Berkeley National
Laboratory (LBL). The HPSS TC is also working with the GridFTP
development project underway at Argonne National Laboratory.

To acquire an HPSS data Grid interface necessary for our test bed, we
requested and received a copy of latest version of GSI-enabled pftp.
The protocol itself is pftp but it supports GSI-enabled AUTH and ADAT ftp-command.

Table~\ref{ftp_prot} shows commands in each
FTP protocol. While GSI-enabled pftp and GridFTP have different command
set for parallel  transfer, buffer management and 
Data Channel Authentication (DCA), the base command set is common.
Fortunately unique functions of each protocol are optional and the 
two protocols are able to communicate.
Installing and testing the GSI-enabled pftp, we proved that the
GSI-enabled pftp daemon from LBL could be successfully accessed from
gsincftp and globus-url-copy with no dcau option(standard globus client utilities). 
From NorduGrid, the server was accessible as well. The below is a
sample XRSL(Extended Resource Specification Language) which utilize 
GSI-enabled pftp server as a storage element(SE) of the NorduGrid.

\vspace{0.5cm}
\begin{minipage}[t]{6cm}
\vspace{0.5cm}
A sample XRSL
\vspace{0.5cm}
\begin{verbatim}
&(executable=gsim1)
(arguments=''-d'')
(inputfiles=
("Bdata.in"
"gsiftp://dt05s.cc:2811/hpss/manabe/data2"))
(stdout=datafiles.out)
(join=true)
(maxcputime="36000")
(middleware="nordugrid")
(jobname="HPSS access test")
(stdlog="grid_debug")%
(ftpThreads=1)
\end{verbatim}
\label{xrsl}
\end{minipage}

\begin{figure}[h]
\includegraphics[width=80mm]{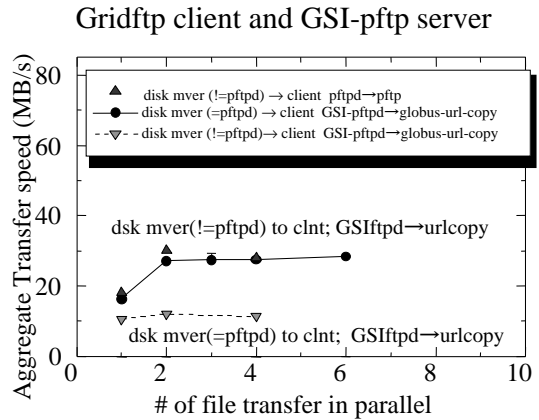}
\caption{Read performance from GSI-enabled pftpd server to Gridftp client}
\label{gridftp}
\end{figure}

In the performance measurement with 2GB file being accessed from
'pftp client',
GSI-enabled pftp server and normal kerberos pftp server had equivalent
elapsed  data transfer time in any situation.
Accessing from `Grid-FTP client', GSI-enabled pftp server and 
normal pftp server, as well, had equivalent transfer time in usual.
However, in the case where multiple disk movers were utilized and 
accessed data and GSI enabled-pftpd server resided in separated disk movers,
transfer speed was halved.
Figure~\ref{gridftp} shows aggregate transfer speed over the
number of independent simultaneous file transfer and shows the 
situation.
After investigating the detailed communication between client and
server, we found the differnece behaviour of the two servers.
In original pftp where 
pftpd running in HPSS core server, data path
is directly established between pftp client and disk mover.
On the other hand, GSI-enabled pftp, data flow was from disk mover,
via pftpd to client host. When the disk mover and pftpd server do not
reside in the same host, two successive network transfer are incurred.

\begin{table}[t]
\begin{center}
\caption{Commands in FTP protocol}
\begin{tabular}{|l|l|}

\hline
  GridFTP     	& GSI-enbled pftp \\
\hline
 SPAS,SPOR,ETET & PBSZ,PCLO,PORPN, 	\\
 ESTO,SBUF,DCAU	& PPOR,PROT,PRTR,PSTO 	\\ 
\hline
 \multicolumn{2}{|c|}{AUTH,ADAT} \\
 \multicolumn{2}{|c|}{RFC959 commands} \\	
\hline
\end{tabular}
\label{ftp_prot}
\end{center}
\end{table}

\section {summary}
ICEPP and KEK configured NorduGrid test bed with HPSS storage server
over high speed GbE WAN. 
Data access performance was measured with several system
configurations in comparison between LAN and WAN access. 
From that, we found that network latency affected  data transfer speed
with HPSS pftp and client API. The ``GSI-enabled pftpd''
developed by LBL was successfully adapted to the interface between
Grid infrastructure and HPSS.

Our paper is a report on work-in-progress. Final results require that
the questions relative to raw TCP performance, server/client protocol
traffic, and pftp a protocol be further evaluated; that any necessary
modifications or parametric changes be acquired form our HPSS team
members; and that measurements be taken again. Further understanding
of the scalability and the limitation of multi-disk mover configurations
would be gained by measuring HPSS network utilization and
performance using higher performance network interfaces adapters,
system software and infrastructure, and processor configurations.


\begin{thebibliography}{9}   % Use for  1-9  references

\bibitem{PBS}
http://www.openpbs.org

\bibitem{netperf}
http://www.netperf.org

\bibitem{gridftp}
http://www.globus.org/datagrid/gridftp.html

\bibitem{hpss}
http://www.sdsc.edu/hpss/

\bibitem{nordugrid}
http://www.nordugrid.org,
You can find NorduGrid papers in this proceedings too.

\bibitem{kekcentral}
S.Yashiro et. al., ``Data transfer using buffered I/O API with HPSS'',
CHEP'01, Beijing, Jul.2001 


\end{thebibliography}
%\begin{thebibliography}{99} % Use for 10-99 references

\end{document}